\documentclass[a4paper,11pt]{article}
\usepackage{pos}

\title{New jet tagging techniques at CMS}

\author*{Dennis Schwarz}

\affiliation{Institut für Experimentalphysik, Universität Hamburg,\\
  Luruper Chaussee 149, Hamburg, Germany}

\emailAdd{dennis.schwarz@desy.de}

\abstract{
The CMS experiment makes use of a large variety of algorithms to identify the origin of particle jets measured in the detector. Through the study of jet substructure properties, jets originating from quarks, gluons, W/Z/Higgs bosons, top quarks and pileup interactions are identified and categorized. We present new techniques based on machine learning approaches developed for the analysis of the data collected during the LHC Run 2 that significantly surpass the performances of classical taggers.
}

\FullConference{%
  40th International Conference on High Energy physics - ICHEP2020\\
  July 28 - August 6, 2020\\
  Prague, Czech Republic (virtual meeting)
}

\begin{document}
\maketitle

\section{Introduction}
The reconstruction and identification of particle jets in the detector are crucial components in many measurements and searches for new phenomena at the LHC. Identification algorithms are developed which rise to the challenge of distinguishing different possible jet origins. In this article, a focus is set on the jet production from hadronic decays of top quarks and heavy bosons at high transverse momenta. In this kinematic regime, the decay products are Lorentz-boosted, become collimated and are reconstructed with a single large-radius jet. In order to identify the origin of those large-radius jets, substructure techniques are indispensable. A classical example of a powerful substructure observable is the jet mass, which is defined as the invariant mass of the sum of all jet constituents. It is sensitive to the mass of the original heavy object once all decay products are reconstructed within the same jet. In order to suppress the Sudakov peak~\cite{substructure} in the jet mass of light quarks and gluon jets, the soft drop algorithm~\cite{substructure,softdrop} is often applied in CMS analyses. With its help, the mass spectrum of jets that include a heavy boson or top quark decay can better be distinguished from gluon or light quark initiated jets. In addition, energy distribution measures can resolve the internal structure of a jet and are used to distinguish the 3-prong decay of top quarks, the 2-prong decays of W, Z and Higgs bosons and the 1-prong structure of jets originating from light quarks and gluons. Taggers that make use of those properties are widely used and have become a standard in CMS analyses. However, recent developments in machine learning approaches made it possible to extract more information from substructure observables and the jet constituents itself and further improve the categorization of jets.

All presented tagging algorithms rely on jets that are clustered from candidates that have been identified by the particle flow algorithm~\cite{particleflow}. A full description of the CMS detector and its subsystems can be found in Ref.~\cite{Chatrchyan:2008aa}.

\section{Jet substructure and machine learning}
Machine learning is an optimal tool for the combination and extraction of features from various jet substructure observables. The boosted event shape tagger~(BEST)~\cite{Sirunyan:2020lcu, CMS-DP-2017-027} is based on substructure observables such as the soft drop jet mass and b tagging information that are fed into a fully connected neural network. In addition, the jet is Lorentz-transformed into its rest frame under the assumption of the mass of the top quark, W, Z or Higgs boson. The angular distribution of decay products is expected to be isotropic and the momenta balanced, if the correct boost into the particle's rest frame is applied. Observables that quantify the isotropic distribution and momentum balance of different boost vectors are fed into the network. Within the neural network, a total of 59 quantities are combined and result in a classification of large-radius jets. The BEST algorithm is capable of identifying multiple jet sources such as hadronic decays of heavy bosons or the top quark. This approach outperforms classical tagging algorithms, which are based on only a few substructure observables.

\begin{figure}
	\centering
	\includegraphics[width=.49\textwidth]{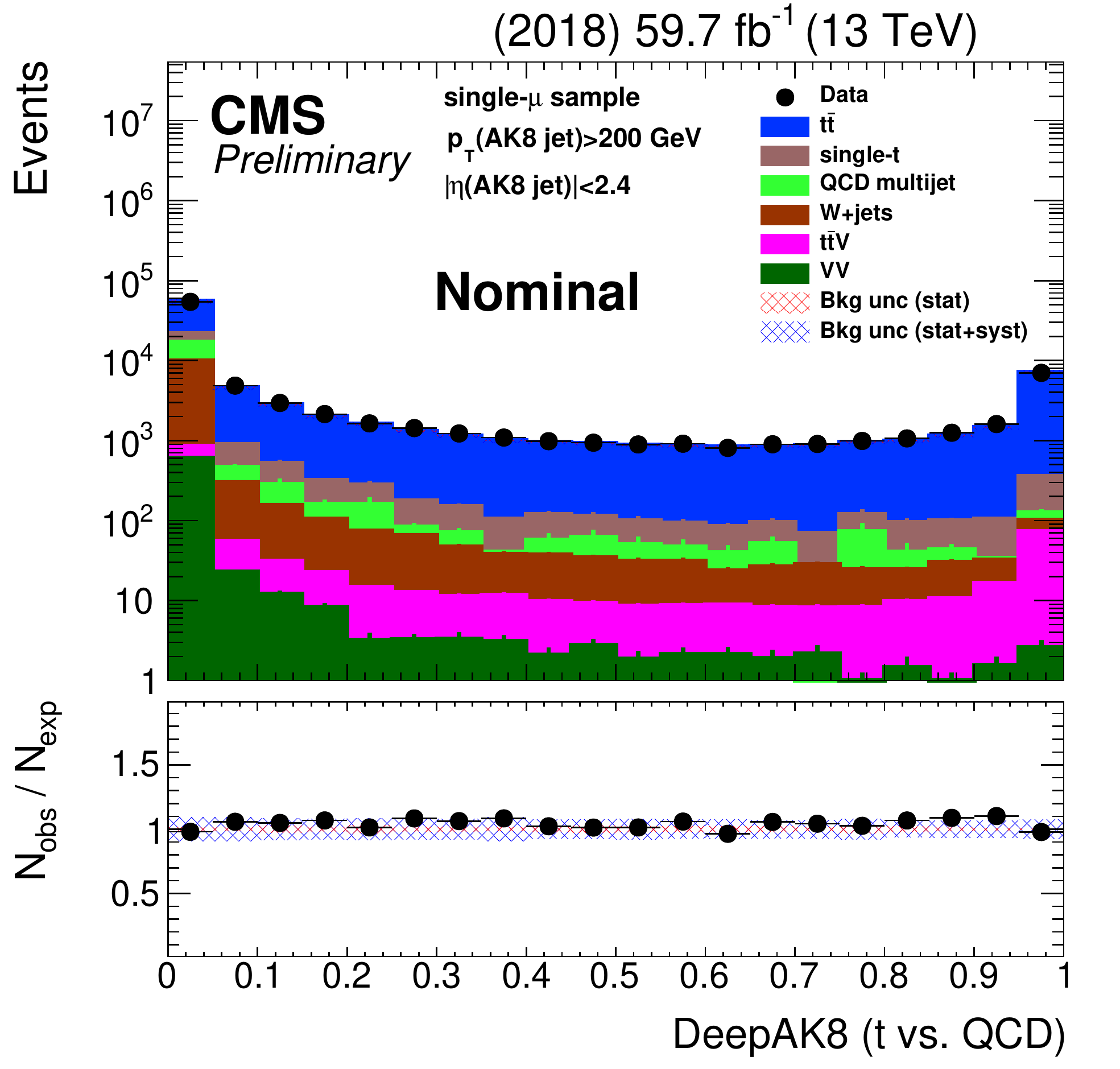}
	\caption{DeepAK8 discriminator distribution for the identification of jets originating from hadronic decays of boosted top quarks. In the upper panel, data~(markers) are compared to simulation~(filled areas). The lower panel shows the ratio and indicates the statistical~(red) and total~(blue) uncertainties as hatched areas. Published in Ref.~\cite{CMS-DP-2020-025}.}
	\label{f:deepak8}
\end{figure}

Since machine learning techniques are very suitable for the distillation of a vast amount of information, algorithms are developed which not only combine jet quantities but properties of the jet constituents itself. The DeepAK8 algorithm~\cite{Sirunyan:2020lcu} is able to combine the properties of up to 100 jet constituents and 7 secondary vertices into its jet categorization. This enormous amount of information is processed in two separate networks that extract features of the constituents and secondary vertices before being combined in one fully connected layer. With the DeepAK8 algorithm, not only the original particle of a decay can be identified but also decay modes can be distinguished. Figure~\ref{f:deepak8} shows the distribution of output values from the node which is used for the separation between light quark and gluon jets and jets that originate from hadronic decays of boosted top quarks. Events are selected that contain a single muon with $p_\text{T} > 55\;\text{GeV}$, at least one b tagged jet and a large-radius jet, clustered with the anti-$k_\text{T}$ algorithm~\cite{antikt} that satisfies $p_\text{T} > 200\;\text{GeV}$. Data are observed to behave very consistently with simulation.

The ParticleNet algorithm~\cite{PhysRevD.101.114025} is based on the same input list of observables as DeepAK8 but uses a different network architecture. Here, a graphic neural network - namely particle clouds - can identify two-particle correlations, which resolve the N-prong structure of a jet.

\section{De-correlation from the jet mass}
\begin{figure}
	\centering
	\includegraphics[width=.49\textwidth]{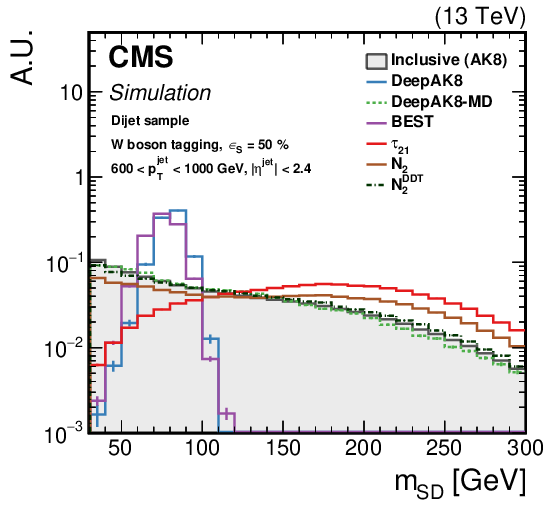}
	\caption{Soft drop jet mass distribution of jets passing the requirements of the nominal and mass de-correlated versions of W tagging algorithms. The distributions are obtained in a dijet sample. Published in Ref.~\cite{Sirunyan:2020lcu}.}
	\label{f:sculpting}
\end{figure}

The jet mass is a key quantity in many taggers and its sensitivity to the jet origin is a central feature. Also machine learning algorithms based on jet constituents are capable of extracting this very observable. With this, jets that pass an identification requirement are more likely to have a mass close to the expected value and the jet mass distribution is sculpt. This introduces difficulties for many analyses that rely on a resonant mass peak in the signal with an otherwise smoothly falling background. Thus, efforts have been made in order to create mass de-correlated tagging algorithms. A mass de-correlated version of the DeepAK8 algorithm, referred to as DeepAK8-MD~\cite{Sirunyan:2020lcu}, has been developed. It makes use of a second neural network that tries to extract and predict the jet mass. The prediction is fed into the original network and a penalty term measuring the accurateness is assigned. This prevents the network from constructing output values which are closely correlated to the jet mass and a smooth jet mass distribution is obtained, as Figure~\ref{f:sculpting} shows.

\section{Expected performance of the algorithms}
\begin{figure}
	\centering
	\includegraphics[width=.49\textwidth]{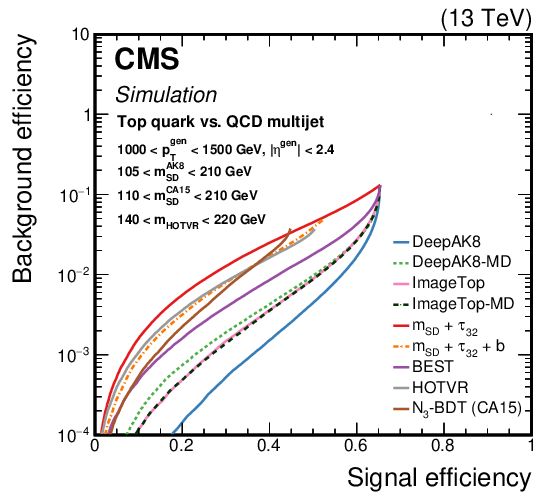}
	\includegraphics[width=.49\textwidth]{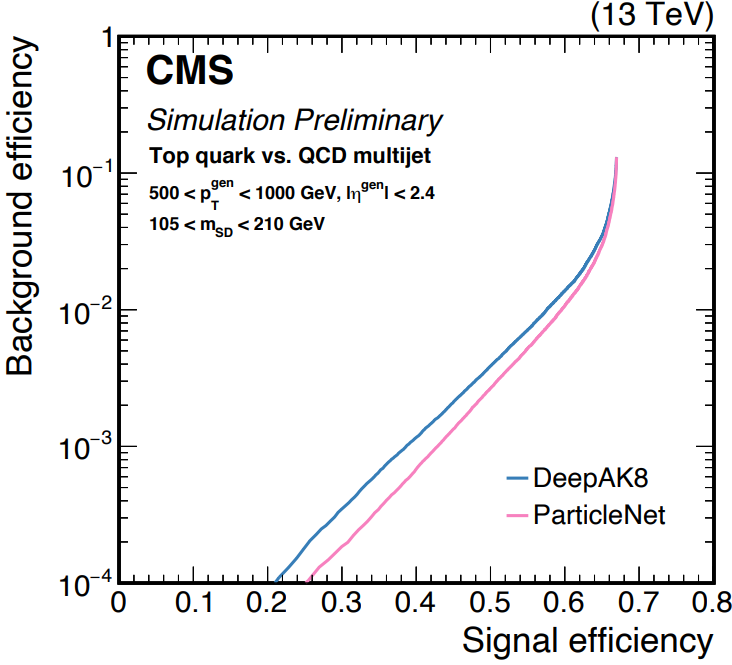}
	\caption{Background efficiency as a function of signal efficiency for a variety of top tagging algorithms. The values are estimated from simulation in a specific kinematic regime. Published in Refs.~\cite{Sirunyan:2020lcu}~(left) and~\cite{PhysRevD.101.114025}~(right).}
	\label{f:validation}
\end{figure}

With all these aforementioned tagging techniques at hand, data analysts have powerful tools, which directly affect and improve a variety of measurements and searches at the LHC. In order to study, compare and understand the multitude of identification algorithms, they are centrally validated within the CMS Collaboration. The measurement of tagging efficiencies is performed using a sample of top quark-antiquark pair events in the lepton+jets channel, where the lepton acts as a tag. Figure~\ref{f:validation} shows scans of the background efficiency as a function of signal efficiency in terms of a receiver operating characteristic (ROC) curve for a variety of top quark identification algorithms measured in simulation. Over the full range, machine learning techniques show a substantial gain in signal efficiency at the same background efficiency and outperform all classical tagging algorithms~($m_\text{SD}+\tau_{32}$, $m_\text{SD}+\tau_{32}+\text{b}$ and HOTVR). In addition, it is shown that ParticleNet achieves an increased tagging efficiency in comparison to DeepAK8, despite relying on the same information. This example shows the potential of new developments in machine learning to not only process large numbers of observables but also extract features in a suitable way.

\section{Validation in data}
\begin{figure}
	\centering
	\includegraphics[width=.6\textwidth]{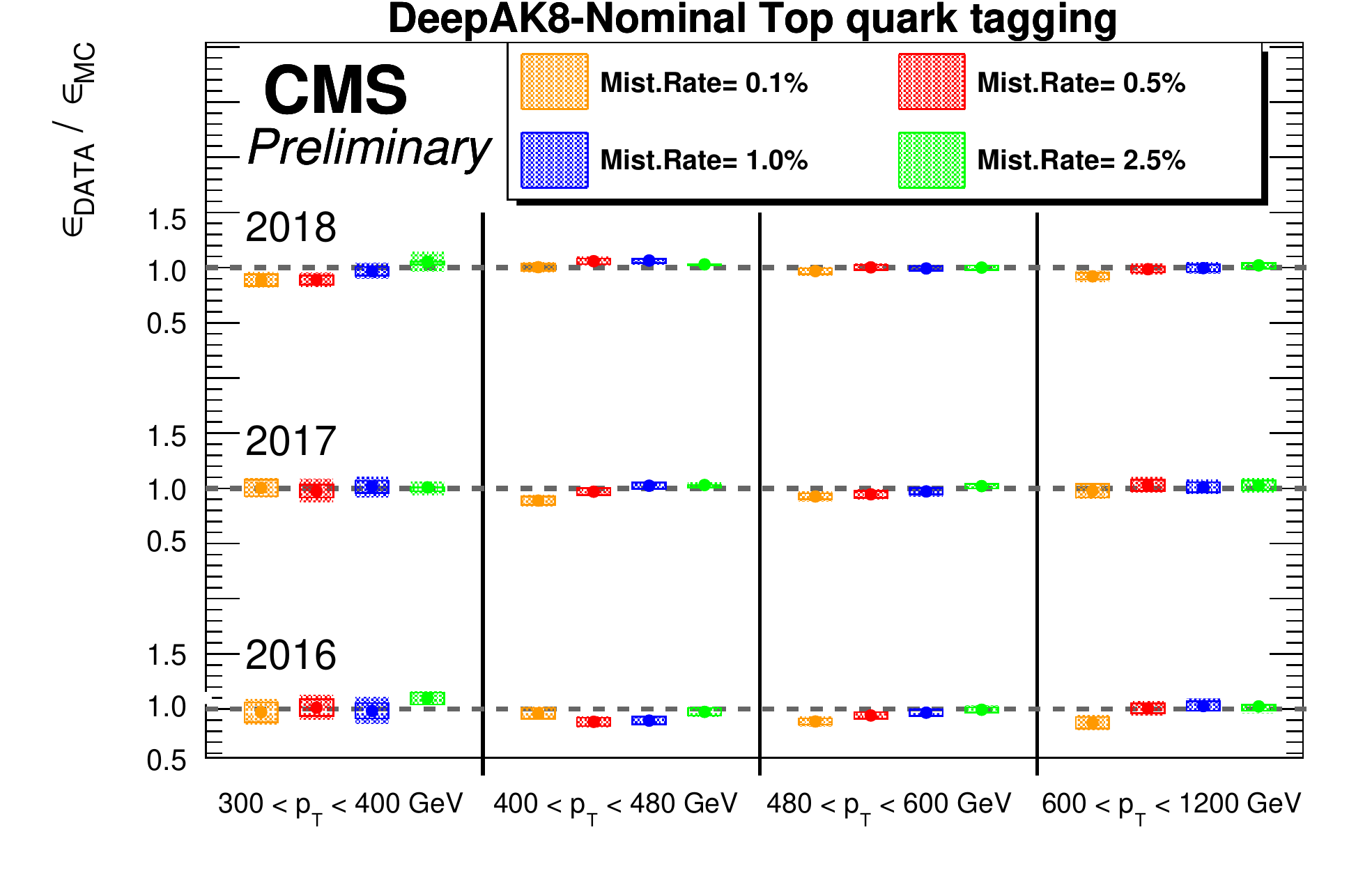} 
	\caption{Correction factors for the DeepAK8 algorithm in the top tagging mode. The factors are separately derived in the 2016, 2017 and 2018 data taking periods of LHC's Run 2 and four different working points in bins of jet momentum. Published in Ref.~\cite{CMS-DP-2020-025}.}
	\label{f:sf}
\end{figure}

By measuring the corresponding efficiencies in data and comparing to those in simulation, correction factors are derived. Those correction factors account for observed differences in jet substructure distributions, which are caused by mismodeling in simulation. Thus, the measurement of tagging efficiencies is not only crucial for all analyses that rely on these tagging techniques but also provide valuable information for our understanding of jet substructure modeling. Figure~\ref{f:sf} shows correction factors, which are estimated for the DeepAK8 algorithm separately for all three data taking periods of LHC's Run 2 and four working points of the tagger. Although machine learning techniques combine a vast amount of substructure information in one output value, all factors are observed to be consistent with unity. Similarly, correction factors are derived for other jet identification algorithms in Refs.~\cite{Sirunyan:2020lcu, CMS-DP-2020-025} and show a similar behavior.

\section{Summary}
The identification of a jet's origin is crucial for many analyses at the LHC. Jet tagging techniques are evolving towards elaborate machine learning approaches which increase the signal efficiency significantly. Data-to-simulation factors are measured in order to correct for deviations in the tagging efficiencies. In addition to dedicated substructure measurements~\cite{substructureCMS, mjetpaper}, these studies are crucial not only for the application of identification techniques but our detailed understanding of jet substructure and its modeling in simulation.

\bibliographystyle{JHEP}
\bibliography{Bibliography}

\end{document}